\documentclass[aps,prl,superscriptaddress,twocolumn,showpacs,amsmath,amssymb]{revtex4}
\usepackage{graphicx}
\usepackage{dcolumn}
\usepackage{bm}

\begin{document}

\title{Activity-dependent stochastic resonance in recurrent neuronal networks}

\author{Vladislav Volman}
\email[Communicating author: ]{volman@salk.edu}%
\affiliation{Center for Theoretical Biological Physics, University of California at San Diego, La Jolla, CA 92093-0319, USA}%
\affiliation{Computational Neurobiology Laboratory, The Salk Institute for Biological Studies, La Jolla, CA 92037, USA}%
\author{Herbert Levine}
\affiliation{Center for Theoretical Biological Physics, University of California at San Diego, La Jolla, CA 92093-0319, USA}%

\begin{abstract}
We use a biophysical model of a local neuronal circuit to study
the implications of synaptic plasticity for the detection of weak
sensory stimuli. Networks with fast plastic coupling show behavior
consistent with stochastic resonance. Addition of an additional
slow coupling that accounts for the asynchronous release of
neurotransmitter results in qualitatively different properties of
signal detection, and also leads to the appearance of transient
post-stimulus bistability. Our results suggest testable hypothesis
with regard to the self-organization and dynamics of local
neuronal circuits.
\end{abstract}

\pacs{Valid PACS appear here}%


\maketitle

\begin{figure}[tbp!]
\centerline{
\resizebox{0.35\textwidth}{!}{\includegraphics{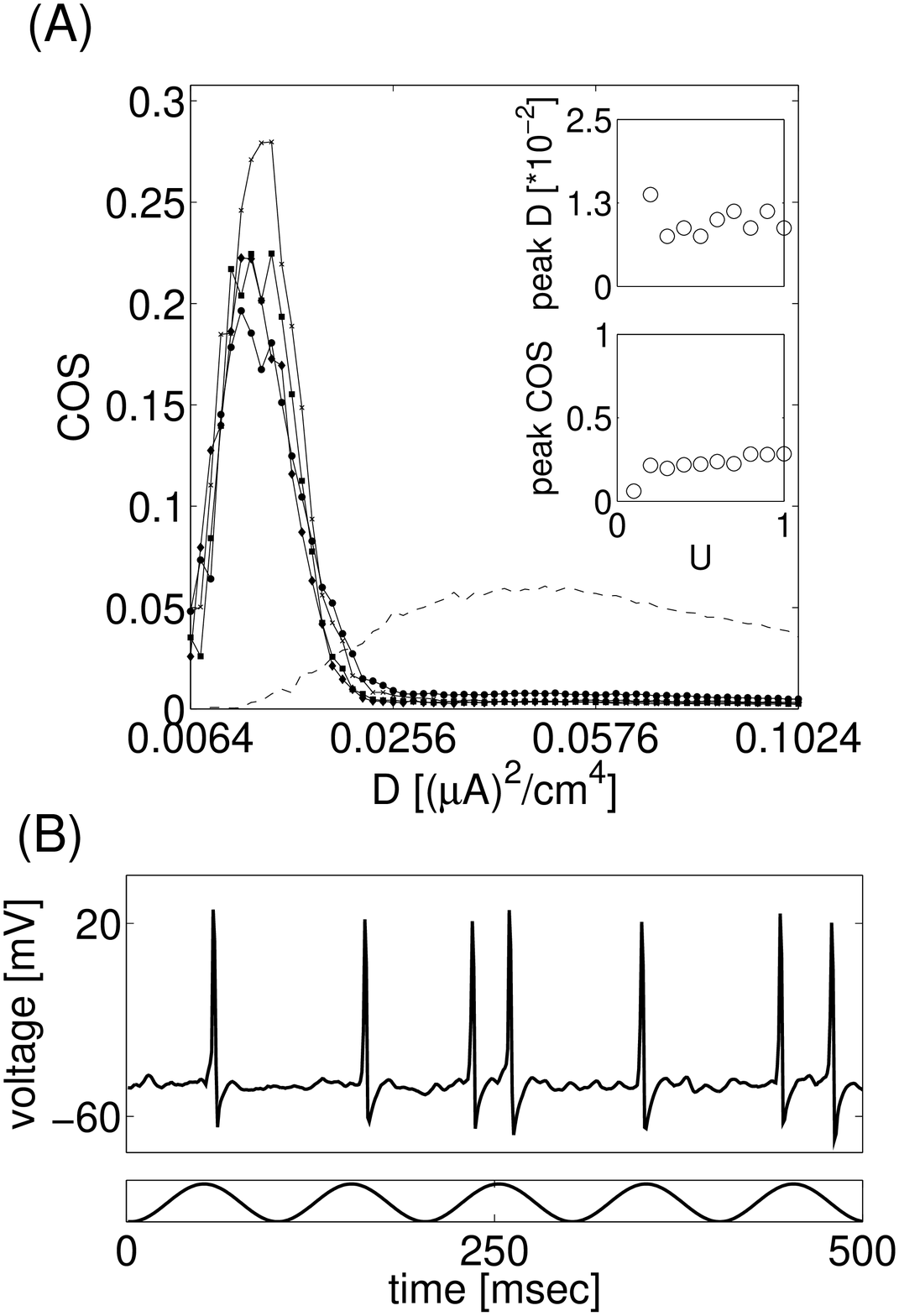}}}
\caption{\label{fig1:Figure1Label}\textbf{Stochastic resonance in
dynamically coupled neuronal networks.} \textit{A)} An uncoupled
cell ensemble exhibits a broad-peak stochastic resonance with
relatively weak coherence of spiking (dashed line). Introduction
of dynamic coupling enables the efficient exchange of
stimulus-related information, and moves the resonance peak to
lower noise intensities. Once above a minimal coupling threshold,
different coupling strengths induce nearly the same
coherence-noise curves (superimposed lines). Both optimal noise
intensity (top inset) and peak coherence (bottom inset) are nearly
independent of synaptic coupling. \textit{B)} Sample neuronal
membrane voltage for $D=1.25\cdot10^{-2},U=0.4,\eta_{max}=0$.}
\end{figure}
\begin{figure}[tbp!]
\centerline{
\resizebox{0.35\textwidth}{!}{\includegraphics{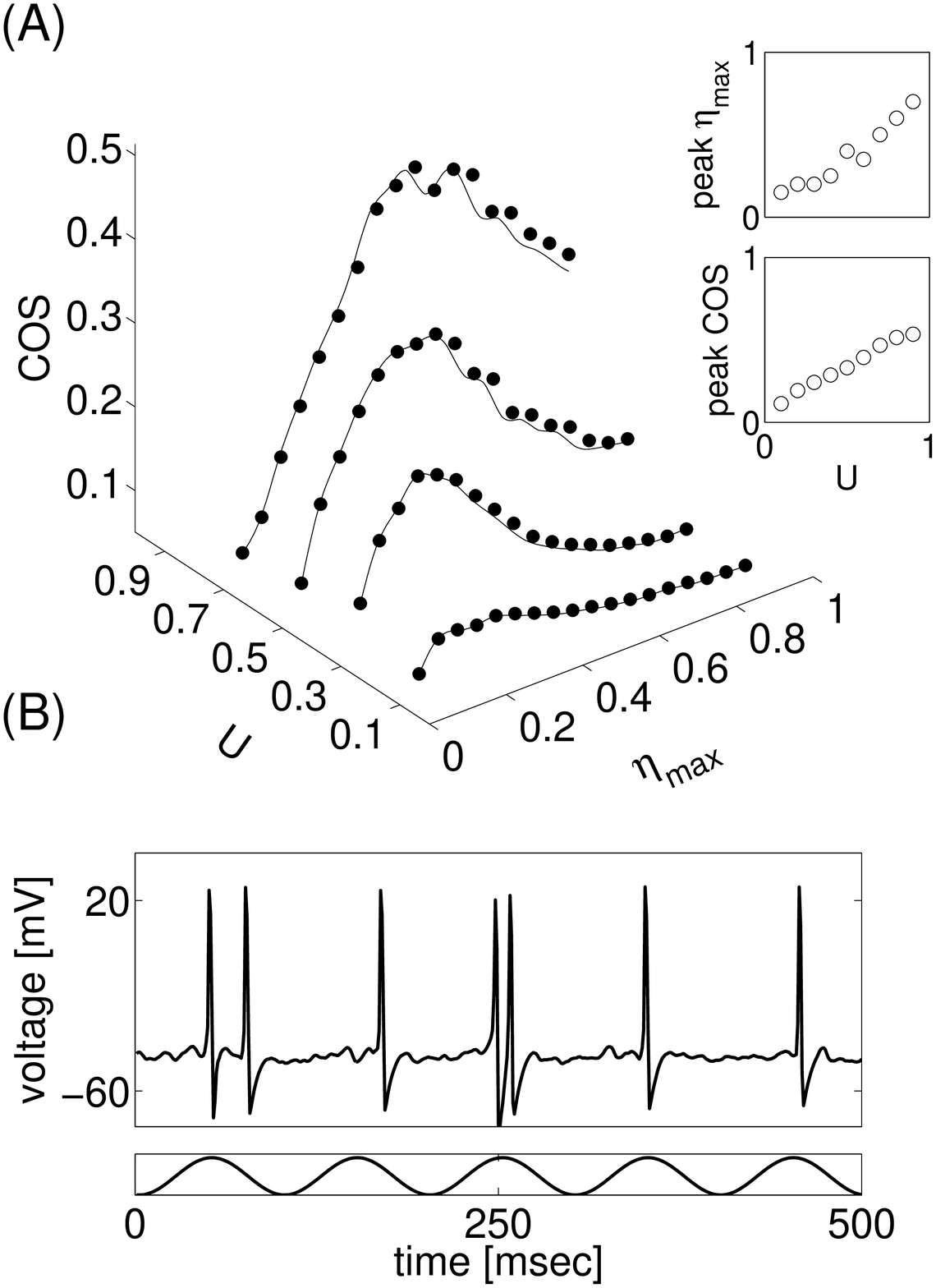}}}
\caption{\label{fig2:Figure2Label}\textbf{Stochastic resonance in
a network with asynchronous release of transmitter.} \textit{A)}
When a slow asynchronous mode of synaptic transmission is
introduced in addition to the fast phasic coupling, the extent of
output spiking coherence depends on the strength of the phasic
coupling ($U$). For clarity of presentation, only the cases
$U=0.1,0.3,0.5,0.7$ are shown. Both the location (top inset) and
magnitude (bottom inset) of the coherence peak are positively
correlated with the strength of evoked synaptic transmission,
underscoring the fact that both kinds of synaptic transmission
share the same pool of synaptic resources. \textit{B)} Sample
neuronal membrane voltage for
$\eta_{max}=0.28,U=0.4,D=0.64\cdot10^{-2}$.}
\end{figure}
\begin{figure}[tbp!]
\centerline{
\resizebox{0.32\textwidth}{!}{\includegraphics{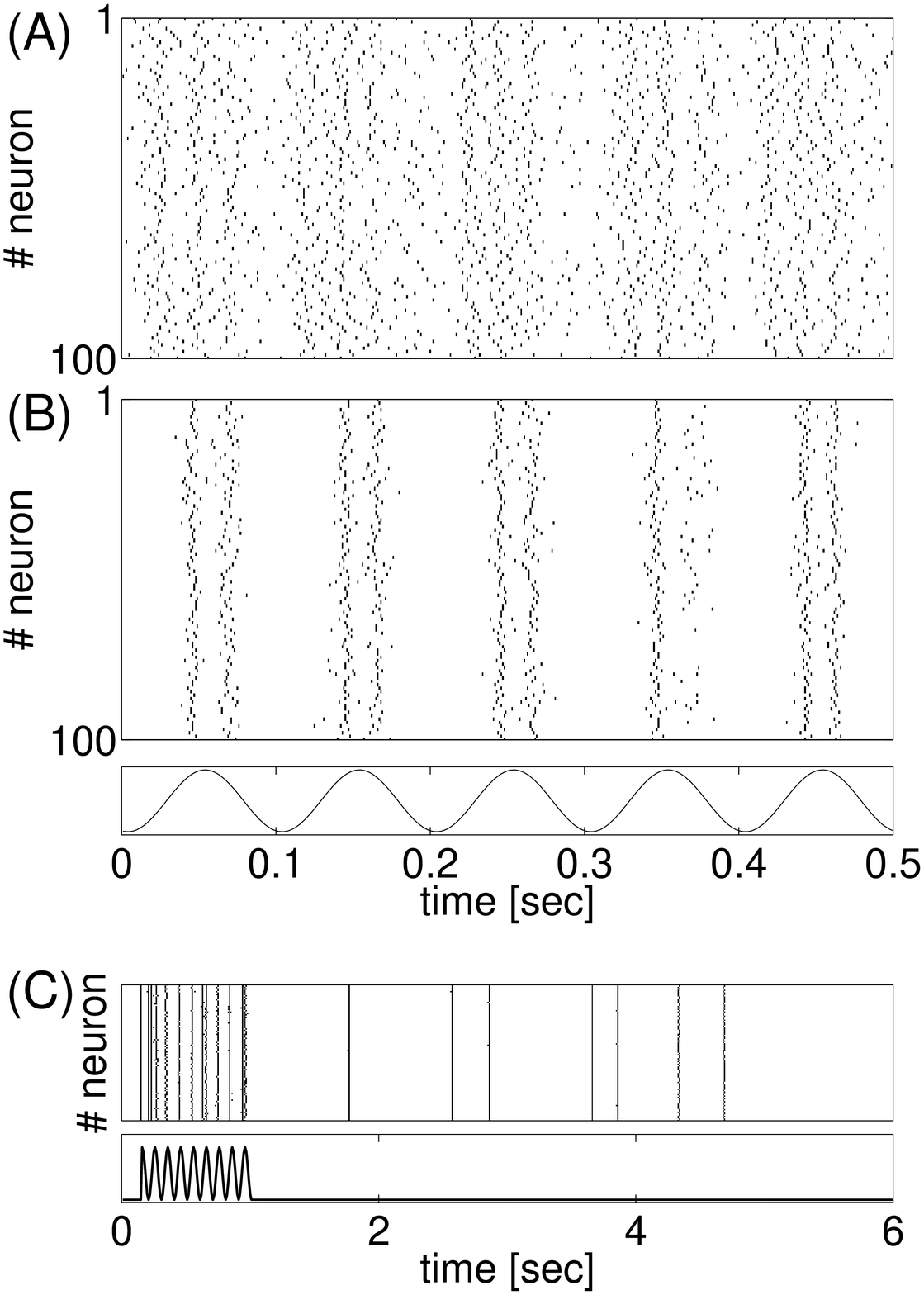}}}
\caption{\label{fig3:Figure3Label}\textbf{AR induces
correlated collective dynamics.} Raster plots show the firing
activity of a neuronal network with $U=0.4$. \textit{A)} A network
with $D=7.84\cdot10^{-2},\eta_{max}=0$ exhibits high-rate,
uncorrelated activity. On the contrary, a network with
$D=0,\eta_{max}=0.8$ \textit{(B)} exhibits burst-like, correlated
collective activity. \textit{C)} A network driven by AR shows transient
bi-stability in its activity (top panel) that persists for seconds
after stimulus removal  (bottom panel).}
\end{figure}
\begin{figure}
\centerline{
\resizebox{0.29\textwidth}{!}{\includegraphics{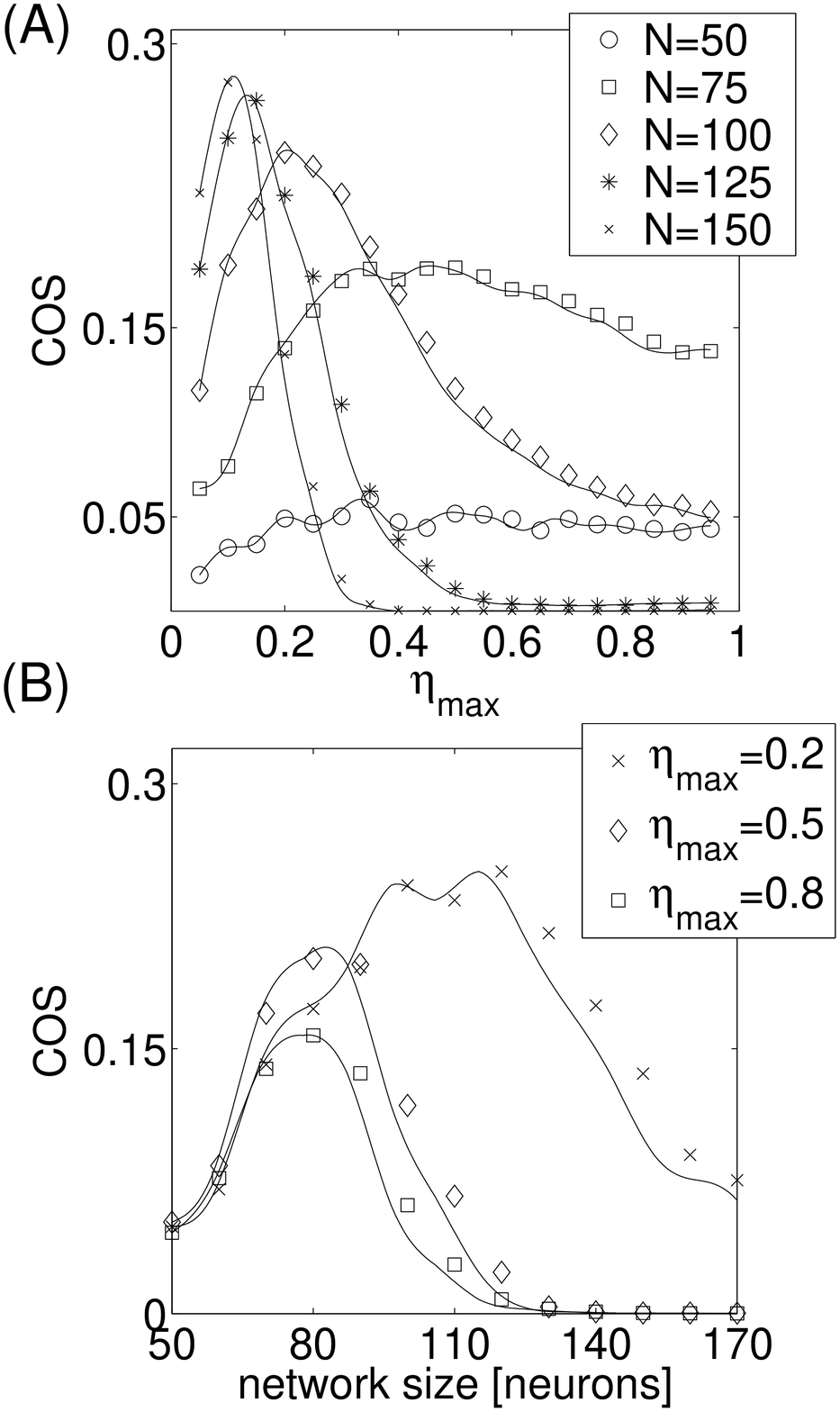}}}
\caption{\label{fig4:Figure4Label}\textbf{System size and
connectivity affect coherence of spiking.} \textit{A)} For a
network with a uniform connection probability $(p=0.1)$, the
profile of coherence as a function of maximal AR rate depends on
the number of neurons. With a $p=const$ scheme, larger networks
lead to higher per cell number of afferents that affects the
effective level of asynchronous release. \textit{B)} Optimal
network size (peaks) that gives rise to maximal coherence depends
on the maximal rate of AR at model synapses. In all cases, the intensity of background synaptic noise is $D=0.64\cdot10^{-2}$.}
\end{figure}

Stochastic resonance (SR) refers to the condition in which noise
and nonlinearity combine together to amplify otherwise
undetectable stimuli \cite{Gammaitoni98}. This simple, yet
important, phenomenon, has received much attention due to its
apparent ubiquity in many nonlinear abiotic \cite{Gammaitoni98}
and biological \cite{SRNeurons} systems. In particular, a number
of studies have raised the possibility that neurons \cite{SRModelNeurons,Rudolph01} and
neuronal cell assemblies \cite{SRNetworks} might utilize SR in order to detect weak
sensory stimuli \cite{SRNeurons}.\\
\indent~For these studies, the noise felt by individual neurons has been assumed to arise from the random summation of a large number of synaptic stimuli~\cite{Rudolph01,Destexhe01}. There is however another important source of noise, that of the stochastic nature of synaptic transmission. In particular, there can occur spontaneous asynchronous release (AR) of neurotransmitter at a rate that is strongly dependent on the pre-synaptic $Ca^{2+}$ concentration and hence strongly dependent on the rate of spike-induced $Ca^{2+}$ intake~\cite{Lau05}. Since a high probability of release can last for
$>0.1~sec$, AR constitutes a challenging example of slow time-scale, activity-dependent noise.\\
\indent~The purpose of this work is to show that SR for {\em local} circuits consisting of roughly 100 neurons (a "micro-column" \cite{Jones00}) coupled via noisy plastic synapses takes a dramatically different form from that seen in investigations to date. As we will see, the coherence of the response continues to depend non-trivially on the coupling strength and the assembly size. Furthermore, the circuit can exhibit short-term memory, by which we mean that spiking will continue to occur for a transient period following removal of the stimulus. These results can be directly tested in experiments on cultured networks \cite{Lau05,Sorkin07} and offer some new insights into the way neuronal systems can be organized for optimal information processing. From the dynamical systems point of view, this work represents a new example of how SR phenomenology can depend on the specific type of noise; this has been considered in only a few examples to date~\cite{DependentNoise}\\
\indent~To proceed, we use a network model that has recently been developed to account for the
occurrence of rhythmic reverberatory responses in
hippocampal cultures \cite{Lau05,Volman07}. The neurons in the network obey Morris-Lecar like
dynamics \cite{MorrisLecar81} with the membrane voltage given by
\begin{equation}
C\dot{V} = -I_{ion}+I_{bg}+I_{syn}+I_{stim}
\label{eq:Eq1_NeuronModel}
\end{equation}
In eqn.\ref{eq:Eq1_NeuronModel}, the ionic current $I_{ion}$
describes the contribution from membrane channels
\cite{NeuronEqs}. The term $I_{bg}$ is a background current that represents summation of a large number of synaptic stimuli from neurons that are not part of the specific local circuit. Rather than explicitly modeling a very large network and imposing a connectivity pattern which embodies the local circuit notion, we instead include these neurons {\em implicitly} by assuming (as in \cite{Destexhe01}) that their contribution is described by a Langevin equation $\dot{I}_{bg}=-I_{bg}/\tau_{n}+\sqrt{D/\tau_{n}}\mathcal{N}(0,1)$, with
$\tau_{n}=10~msec$ and $\mathcal{N}(0,1)$ being uncorrelated Gaussian noise with zero mean and unitary variance. The synaptic current due to the local circuit is modeled as $I_{i}^{syn}=-\Sigma\bar{g}Y_{ij}(t)V_{i}$, with $\bar{g}\in
[0.5,0.8]~mS/cm^{2}$ being the maximal value of synaptic
conductance, the sum running over the set of input channels, and
the term $Y_{ij}$ as described below. With the parameters as given
in \cite{NeuronEqs}, the transition from quiescence to
regular spiking occurs through a Hopf bifurcation.\\
\indent~To capture the
dynamical aspects of synaptic coupling, we assume that at any
time, presynaptic resources can be in a recovered state (described by
the state variable $X$ in equations below), in an active state
(described by the state variable $Y$), or in an inactive state
(described as $Z=1-X-Y$) \cite{Tsodyks00}. The dynamics for the
$j\rightarrow i$ presynaptic terminal are
\begin{subequations}
\label{eq:Eq_SynModel}
\begin{eqnarray}
\dot{X}_{ij} &=& \frac{Z_{ij}}{\tau_{r}}-X_{ij}(U\delta(t-t_{s}^{j})+\xi\delta(t-t_{a}^{j})) \\
\dot{Y}_{ij} &=& \frac{-Y_{ij}}{\tau_{d}}+X_{ij}(U\delta(t-t_{s}^{j})+\xi\delta(t-t_{a}^{j})) \\
\dot{Z}_{ij} &=& \frac{Y_{ij}}{\tau_{d}}-\frac{Z_{ij}}{\tau_{r}} \\
\eta(c) &=& \eta_{max}\frac{c^4}{c^4+K_{a}^{4}} \\
\dot{c} &=& \frac{-\beta c^{2}}{c^{2}+K_{Ca}^{2}}+\gamma
log(\frac{c_{o}}{c})\delta(t-t_{s}^{j})+I_{p}
\end{eqnarray}
\end{subequations}
\indent~At each presynaptic terminal of the $i$-th neuron, the
fraction of active resources $(Y_{ij})$ experiences a brief
increase of magnitude $UX_{ij}$ when, at time $t_{s}^{j}$, an
action potential from $j$-th neuron invades the presynaptic
terminal. Alternatively, a relatively small amount of resource can
be maintained in an active state by the asynchronous release of
synaptic resource that occurs at times $t_{a}^{j}$ with
$Ca^{2+}$-dependent rate $\eta(c)$ and amplitude $\xi$. The rate
of asynchronous release (probability to observe an event during
the interval $[t_{a},t_{a}+dt]$, modeled as Poisson process) is taken to be a Hill function of the
presynaptic residual $Ca^{2+}$ concentration, $c$
\cite{Volman07,Ravin97}. This residual $Ca^{2+}$ accumulates at
presynaptic terminals in an activity-dependent way that is
proportional to electrochemical gradient across the membrane, and
is extruded into the extra-cellular space by a non-linear pump.
The term $I_{p}$ ensures that the minimal $Ca^{2+}$ concentration
is $\approx 60~nM$. Parameters are given in \cite{SynParams}. Note
that the phasic $(UX_{ij})$ and asynchronous $(\xi X_{ij})$ terms
are both proportional to the amount of available resource,
$X_{ij}$, underscoring the activity-dependent competition between
these two different coupling modes \cite{ARCompetition}.\\
\indent~To assess the extent to which an individual neuron and/or
a network can exhibit coherent activity, we use the coherence of
spiking (COS) measure
\cite{COSRef,Rudolph01}. Given a weak external sub-threshold stimulation of period
$T$, $I_{stim}(t)=1\frac{nA}{cm^{2}}sin(2\pi\frac{t}{T})$, the COS measure is defined here as
$C_{S}\equiv\frac{N(0.9T<=ISI<=1.1T)}{N(ISI)}$, that is, the
fraction of inter-spike-intervals (ISIs) that are within $20\%$ of
stimulus period, $T=0.1~sec$. All results, unless otherwise
indicated, are for a network of $N=100$ neurons that have
probability $p=0.1$ to establish connections with their peers.\\
\indent~We first analyze the response of an uncoupled neuronal
network to weak sub-threshold periodic stimulation and different
(controlled) intensities of synaptic background noise, $I_{bg}$. In agreement with previous studies
\cite{SRModelNeurons}, we find that there exists an optimal level
of noise for which a model neuron exhibits a maximal coherence of
spiking (Fig.\ref{fig1:Figure1Label}A, dashed line). Coupling the
model neurons by activity-dependent synapses (as in eqs.
\ref{eq:Eq_SynModel}) while setting $\eta_{max}=0$ (no
asynchronous release) moves the resonant peak towards lower noise
intensities. As Fig.\ref{fig1:Figure1Label}A (insets) shows, the
location and the height of the new peak is largely independent of
the coupling parameter, $U$. This observation is consistent with
the notion of efficient signal propagation on random graphs - once
$U$ is above critical coupling threshold, an SR-like activation of
one neuron will quickly spread the word to other neurons,
regardless of the exact value of $U$.\\
%
\indent~Introduction of activity-dependent asynchronous release of
neurotransmitter results in a qualitatively different picture.
The coherence measure as a function of evoked and asynchronous
release is shown in Fig.\ref{fig2:Figure2Label}A. It is clear that
the spiking coherence increases significantly for higher values of the resource
utilization parameter $U$. The optimal level of AR needed to produce maximal coherence (peaks in
Fig.\ref{fig2:Figure2Label}A) also depends on the value of  $U$. Stronger evoked transmission
will quickly deplete the available resources; therefore, since
asynchronous and evoked releases draw from the same pool of
vesicles, higher rate of spontaneous events is needed to achieve
significant spiking coherence (top inset of
Fig.\ref{fig2:Figure2Label}A). For higher values of $\eta_{max}$,
when the combined action of AR and $I_{bg}$ masks the stimulus by
making the cell spike more frequently, the coherence measure
converges to low values. On the other hand, strong coupling and
fast depletion of resources provide a constraint for spiking
activity, resulting in higher overall coherence for
higher resource usage (bottom inset of Fig.\ref{fig2:Figure2Label}A).\\
\indent~The distinctive effect of AR (as compared with $I_{bg}$)
is further assessed by analyzing the collective dynamics for high
$\eta_{max}$ (vs. high $D$). Subjecting the network to
high-intensity dynamics-independent noise
(Fig.\ref{fig3:Figure3Label}A) results in high-rate, weakly
correlated, activity. On the other hand, as
Fig.\ref{fig3:Figure3Label}B shows, the combined action of strong
AR and synaptic depression significantly sharpens the network's
response to the stimulus. Further, the prolonged time-scale of AR
enables the network to "remember" the stimulus seconds after its
cessation (Fig.\ref{fig3:Figure3Label}C).\\
\indent~The observation that coherence of spiking depends on the
strength of dynamic coupling prompted us to explore how networks'
parameters affect its ability to detect weak stimuli. To this end,
we considered the performance of different size networks, for a
range of AR rates. For easier interpretation of results, we assume
here that, for all cases, $U=0.3$. Figure \ref{fig4:Figure4Label}A shows that the profiles of
COS curves are different for different network sizes. Due to the
$p=const$ constraint, neurons in larger networks are subject to
higher levels of asynchronous release in their inputs; as a
result, the resonant peak moves toward lower values of $\eta_{max}$.
Conversely, fixing the value of $\eta_{max}$ and plotting the COS
measure as a function of network size (as is in
Fig.\ref{fig4:Figure4Label}B) reveals that the optimal network size
(giving maximal coherence) depends on the level of AR at
individual model synapses. Thus, in a network with plastic
coupling, synaptic parameters might provide constraints for the
sizes of cell assembly.\\
Stochastic resonance relies on the cooperativity between
noise, nonlinearity and a weak sub-threshold stimulus
\cite{Gammaitoni98}. In most examples,  the noise
is taken to be independent of the characteristics of the weak sub-threshold stimulus (but see \cite{DependentNoise}).
Here, we have investigated the properties of signal processing in
local recurrent neuronal networks with plastic coupling and asynchronous
release of neurotransmitter, where the noise is inherently coupled to the signal. We found that in plastic networks without AR, the characteristics of stochastic resonance (location
and height of peak coherence) only weakly depend on the strength
of synaptic coupling. On the contrary, introduction of AR  leads to a strong dependence of SR
properties on network parameters.\\
\indent~These observations suggest that
asynchronous release of neurotransmitter might play an important
role in neuronal dynamics~\cite{ARSignificance}. Information that is contained in weak signals should not
only be detected and amplified by brain circuitry; a network has
to have the ability to transiently "hold" knowledge about
preceding events. As shown in \cite{Lau05,Volman07}, a brief
stimulus delivered to the network evokes reverberatory
activity that is sustained by the asynchronous release of
neurotransmitter and lasts for several seconds. Our results (Fig.\ref{fig3:Figure3Label}C) together with experimental
observations \cite{Lau05} and prior modeling \cite{Volman07},
suggest that AR can be instrumental in detection, amplification,
and transient holding of weak sensory
stimuli.\\
\indent~This study leads to several potentially interesting
conclusions. First, we showed here that the ability of a neuron
(that is embedded in a neuronal network) to detect and amplify weak
stimuli might crucially depend on the form of feedback from the
network, and in particular on the plasticity features of the effective
connectivity. Second, our results suggest that the plasticity of
synaptic connections might provide an important constraint for the
optimal number of neurons in a local circuit. With this perspective, the local network with strong inter-connectivity is optimized for signal detection, with distant neurons providing contextual information in  the form of an overall background noise signal. Cultured networks can
provide an adequate framework to test the validity of our
conclusions. State of the art techniques allow one to grow small
networks of controlled size, geometry and connectivity
\cite{Sorkin07}. Future experiments will determine how these
parameters affect the ability of a network to process weak stimuli.\\
\indent~We thank W.J. Rappel and T.J. Sejnowski for stimulating
discussions. This research has been supported by the NSF-sponsored
Center for Theoretical Biological Physics (grant~nos.~PHY-0216576
and PHY-0225630).


\begin{thebibliography}{17}
\bibitem{NeuronEqs}
For each model neuron, the ionic current was
$I_{ion}=G_{Na}m_{\infty}(V-E_{Na})+G_{K}w(V)(V-E_{K})+G_{L}(V-E_{L})$.
The fraction of open $K^{+}$ channels evolved as
$dw/dt=0.15(w_{\infty}(V)-w(V))cosh((V-V_{3})/2V_{4})$. The
steady-state fraction of $Na^{+}$ and $K^{+}$ channels were,
correspondingly, $m_{\infty}=0.5(1+tanh((V-V_{1})/V_{2}))$ and
$w_{\infty}=0.5(1+tanh((V-V_{3})/V_{4}))$. Values of parameters
were: $E_{Na}=50~mV, E_{K}=-100~mV, E_{L}=-55.8~mV, V_{1}=-1.2~mV,
V_{2}=23~mV, V_{3}=-2~mV, V_{4}=21~mV, G_{Na}=11~mS/cm^{2},
G_{K}=10~mS/cm^{2}, G_{L}=1.5~mS/cm^{2}, C=1~\mu~F/cm^{2}$. Equations were solved
using second-order Runge-Kutta method with $dt=0.1~msec$.
\bibitem{SynParams}
The parameters used to model synaptic transmission were
$\tau_{d}=5~msec, \tau_{r}=0.6~sec, K_{a}=0.1~\mu M,
K_{Ca}=0.4~\mu M, \beta=2~\mu M/sec, \gamma=80~nM, C_{0}=2~mM,
I_{p}=0.11~\mu M/sec, \xi=10^{-3}$
\end{thebibliography}


\end{document}